\documentclass[prb,twocolumn,aps,showpacs,floatfix,superscriptaddress]{revtex4}

\usepackage{graphicx}% Include figure files
\usepackage{dcolumn}% Align table columns on decimal point
\usepackage{bm}% bold math
\usepackage{amsmath}

\begin{document}

\title{Aharonov-Bohm effect and broken valley-degeneracy in graphene rings}
\author{P. Recher}
\affiliation{Instituut-Lorentz, Universiteit Leiden, P.O. Box 9506, 2300 RA Leiden, The Netherlands}
\affiliation{Kavli Institute of Nanoscience, Delft University of Technology, Lorentzweg 1, 2628 CJ Delft, The Netherlands}
\author{B. Trauzettel}
\affiliation{Department of Physics and Astronomy, University of Basel, Klingelbergstrasse 82, 4056 Basel, Switzerland}
\author{A. Rycerz}
\affiliation{Marian Smoluchowski Institute of Physics, Jagiellonian University, Reymonta 4, 30-059 Krak$\acute{o}$w, Poland}
\author{Ya. M. Blanter}
\affiliation{Kavli Institute of Nanoscience, Delft University of Technology, Lorentzweg 1, 2628 CJ Delft, The Netherlands}
\author{C. W. J. Beenakker}
\affiliation{Instituut-Lorentz, Universiteit Leiden, P.O. Box 9506, 2300 RA Leiden, The Netherlands}
\author{A. F. Morpurgo}
\affiliation{Kavli Institute of Nanoscience, Delft University of Technology, Lorentzweg 1, 2628 CJ Delft, The Netherlands}

\begin{abstract}
We analyze theoretically the electronic properties of Aharonov-Bohm rings made
of graphene. We show that the combined effect of the ring confinement and
applied magnetic flux offers a controllable way to lift the orbital degeneracy
originating from the two valleys, even in the absence of intervalley
scattering. The phenomenon has observable consequences on the persistent
current circulating around the closed graphene ring, as well as on the ring
conductance. We explicitly confirm this prediction analytically for a circular
ring with a smooth boundary modelled by a space-dependent mass term in the
Dirac equation. This model describes rings with zero or weak intervalley
scattering so that the valley isospin is a good quantum number.  
The tunable breaking of the valley degeneracy by the flux allows for the
controlled manipulation of valley isospins.  
We compare our analytical model to another type of ring with strong intervalley
scattering. For the latter case, we study a ring of hexagonal form with
lattice-terminated zigzag edges numerically.  
We find for the hexagonal ring that the orbital degeneracy can still be
controlled via the flux, similar to the ring with the mass confinement.
\end{abstract}

\date{\today}
\pacs{73.23.-b, 73.23.Hk, 73.23.Ra, 81.05.Uw}
\narrowtext \maketitle

\section{Introduction}
Graphene offers the remarkable possibility to probe predictions of quantum
field theory in condensed matter systems, as its low-energy spectrum is
described by the Dirac-Weyl Hamiltonian of massless fermions.\cite{Geimreview}
However, in graphene, Dirac electrons occur in two degenerate families, corresponding to the presence of two different valleys in the band structure -- a phenomenon known as
``fermion doubling''. This valley degeneracy makes it difficult to observe the {\it intrinsic} physics of a single valley in experiments, because in many cases the contribution of one valley to a measurable quantity is exactly cancelled by the contribution of the second valley. A prominent example that single-valley physics is interesting, is the
production of a fictitious magnetic field in a single valley by a
lattice defect or distortion.\cite{AMFG} The field has the opposite sign in the other valley, so its effect is hidden when both
valleys are equally populated. Another example is the existence of weak antilocalization in
diffusive graphene, which is destroyed by intervalley scattering.\cite{WAL} Therefore, from a fundamental point of view, it is desirable to
find a feasible and controlled way to lift the valley degeneracy in graphene. From a more practical
point of view, the lifting of the orbital degeneracy is essential for
spin-based quantum computing in graphene quantum dots
\cite{graphenespinqubits}, which is a promising direction of future research
because of the superior spin coherence properties expected in
carbon structures.
\begin{figure}
\begin{center}
\hbox{\qquad\qquad\qquad\,\,\resizebox{3.8cm}{!}{\includegraphics{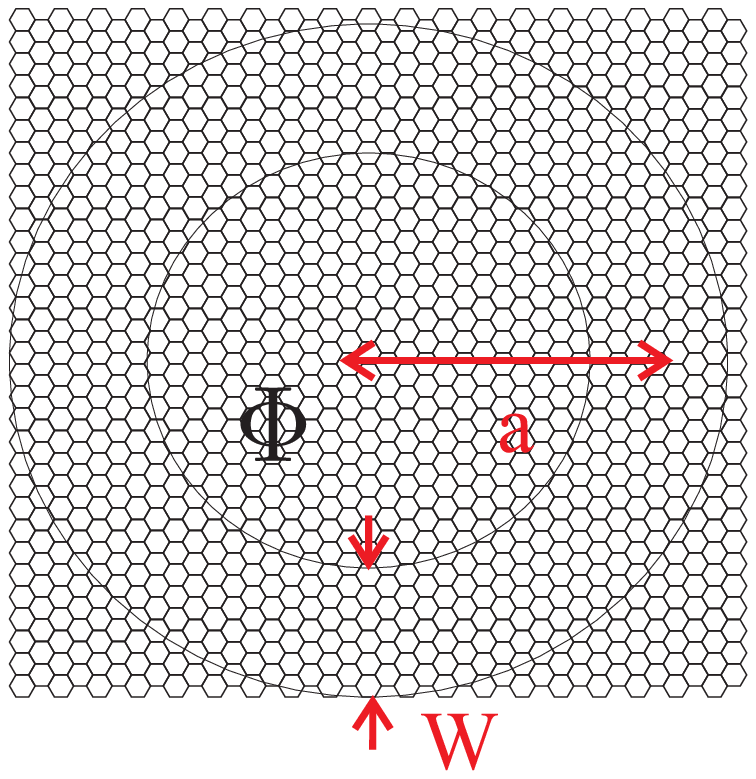}}}
\end{center}
\vspace{-0.8cm} \caption{A circular graphene ring of radius $a$ and width $W$ subjected
  to a magnetic flux $\Phi$ threading the ring.}
\end{figure}

Here, we show that the confinement of electrons in graphene in an Aharonov-Bohm (AB) ring (see Fig.~1) provides a conceptually simple way to achieve a controlled lifting of the valley degeneracy. We find that the ring
confinement -- which breaks the effective time reversal symmetry (TRS) in a
single valley in the absence of intervalley scattering \cite{Berry,Antonremark} -- leads generically to a lifting of the valley degeneracy controllable by magnetic flux. 
To demonstrate this, we choose an analytical model with a smooth ring boundary
described by a mass term in Section II. Such a mass term might be generated in
a real system by the influence of the lattice of the substrate on the band
structure of the sample, see Refs.~\onlinecite{Giova07,Zhou07} for concrete examples of such
an effect.
However, our conclusions are not restricted to the mass confinement but potentially hold for any boundary that preserves the valley isospin as implied by general symmetry arguments.\cite{Antonremark} We show that the signature of the broken valley
degeneracy is clearly visible in the persistent current and the conductance
through the ring. It is further illustrated how to use the lifting of the valley degeneracy with flux to {\it manipulate} and {\it measure} the valley isospin. 
 
In Section III, we compare our analytical model for a smooth boundary to a system where intervalley scattering is strong. For this purpose, we study a ring of hexagonal shape with zigzag edges where intervalley scattering is induced at the corners of the hexagon. We calculate the spectrum numerically in a tight-binding approach and find that the orbital degeneracy can still be tuned by the magnetic flux, similar to the analytical model. We test this ability against a small distortion of the 6-fold symmetry of the ring and find small avoided crossings at zero flux.

\section{Ring with smooth boundary}
In this section, we analyze in detail the spectral properties of a graphene ring subjected to a magnetic flux and its signatures in persistent current and conductance through the ring assuming a smooth confinement induced by a space-dependent mass term in the Dirac equation. We also discuss how to address the valley degree of freedom in such a ring structure. 
\subsection{Spectrum}
The graphene ring with valley degree of freedom $\tau=\pm$ is modeled by the Hamiltonian ($\hbar=c=1$)
\begin{equation}
\label{E1}
H_{\tau}=H_{0}+\tau V(r){\sigma}_{z},
\end{equation}
where we use the valley isotropic form \cite{AKBE} for $H_{0}=v({\bf p}+e{\bf A})\cdot{\bm \sigma}$ with ${\bf p}=-i\partial/\partial {\bf
r}$, $-e$ being the electron charge,  $v$ the Fermi velocity and
 ${\sigma}_{i=x,y,z}$ are the Pauli matrices. The vector potential is ${\bf A}=(\Phi/2\pi r){\bf e}_{\varphi}$ with $\Phi$ the magnetic flux threading the ring, see Fig.~1. The
term proportional to ${\bf \sigma}_{z}$ in Eq.~(\ref{E1}) is a
mass term confining the Dirac electrons on the ring. Introducing
polar coordinates, the Hamiltonian $H_{0}$ is written as
\begin{multline}
\label{E2}
H_{0}(r,\varphi)=-iv(\cos\varphi\sigma_{x}+\sin\varphi\sigma_{y})\partial_{r}\\
-iv(\cos\varphi\sigma_{y}-\sin\varphi\sigma_{x})\frac{1}{r}\left(\partial_{\varphi}+i\frac{\Phi}{\Phi_{0}}\right).
\end{multline}
The angular orbital momentum in the $z$-direction is
$l_{z}=-i\partial_{\varphi}$ and $\Phi_{0}=2\pi/e$.
The two valleys $\tau=\pm$ decouple and we can solve the spectrum for each valley separately, $H_{\tau}\psi_{\tau}=E\psi_{\tau}$. We note the similarity of $H_{\tau}$ to a ring with the Rashba interaction.\cite{Meijer,Frustaglia} However, an important difference is that for graphene, the confinement potential acts on the (pseudo)spin, whereas for Rashba interaction the confinement potential is spin-independent. As a consequence, the confining potential in Eq.~(\ref{E1}) couples the pseudo-spin components and breaks effective TRS (${\bm p}\rightarrow -{\bm p}$, ${\bm \sigma}\rightarrow -{\bm \sigma}$) even in the absence of a flux $\Phi$.\cite{Berry}
Since $H_{\tau}$ commutes with $J_{z}=l_{z}+\frac{1}{2}\sigma_{z}$, its
eigenspinors $\psi_{\tau}$ are
eigenstates of  $J_{z}$,
\begin{equation}
\label{E3}
\psi_{\tau}(r,\varphi)=e^{i(m-1/2)\varphi}\left(\begin{array}{l}
   \chi_{1}^{\tau}(r)    \\
       \chi_{2}^{\tau}(r) e^{i\varphi}
\end{array}\right)
\end{equation}
with
eigenvalues $m$, where $m$ is a half-odd integer,
$m=\pm\frac{1}{2}, \pm\frac{3}{2}, \dots$
 The radial component  $\chi_{\tau}(r)\equiv
 (\chi_{1}^{\tau}(r),\chi_{2}^{\tau}(r))$
satisfies ${\widetilde H}_{\tau}(r)\chi_{\tau}(r)=E\chi_{\tau}(r)$ with
\begin{equation}
\label{E5} {\widetilde
H}_{\tau}(r)=-iv\sigma_{x}\partial_{r}+\tau V(r)\sigma_{z}+v\sigma_{y}\frac{1}{r}\left(\begin{array}{cc}{\overline m}-\frac{1}{2} & 0\\ 0&\overline {m}+\frac{1}{2}\end{array}\right),
\end{equation}
where we have defined $\overline{m}=m+(\Phi/\Phi_{0})$.
\begin{figure}
\begin{center}
\hbox{\quad\resizebox{7.2cm}{!}{\includegraphics{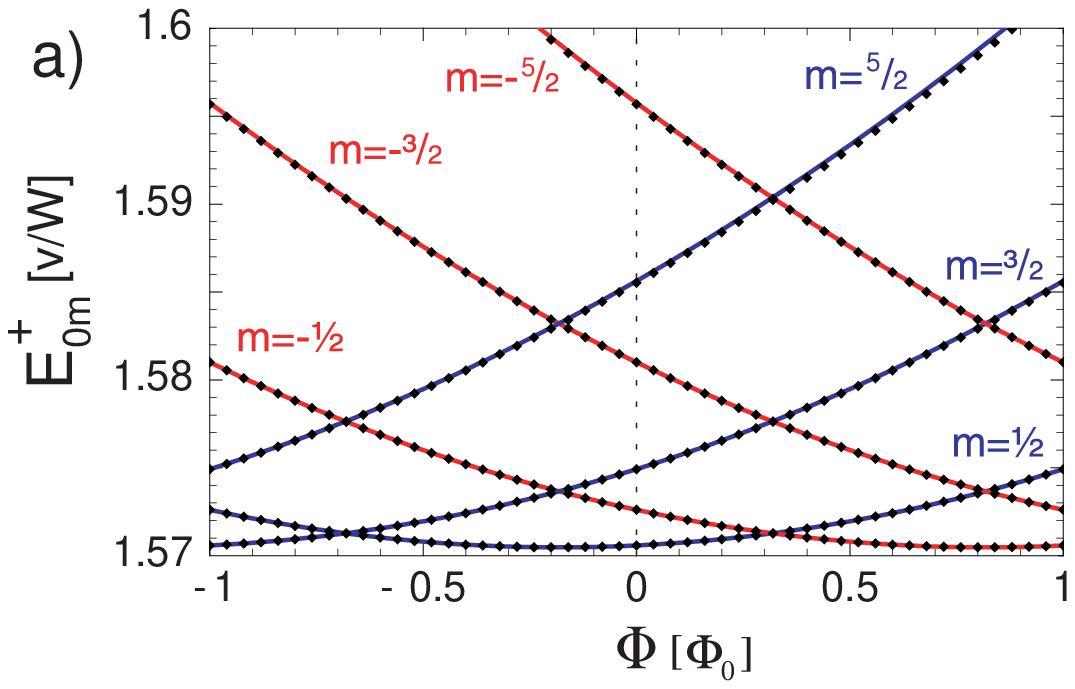}}}
\hbox{\quad\resizebox{7.2cm}{!}{\includegraphics{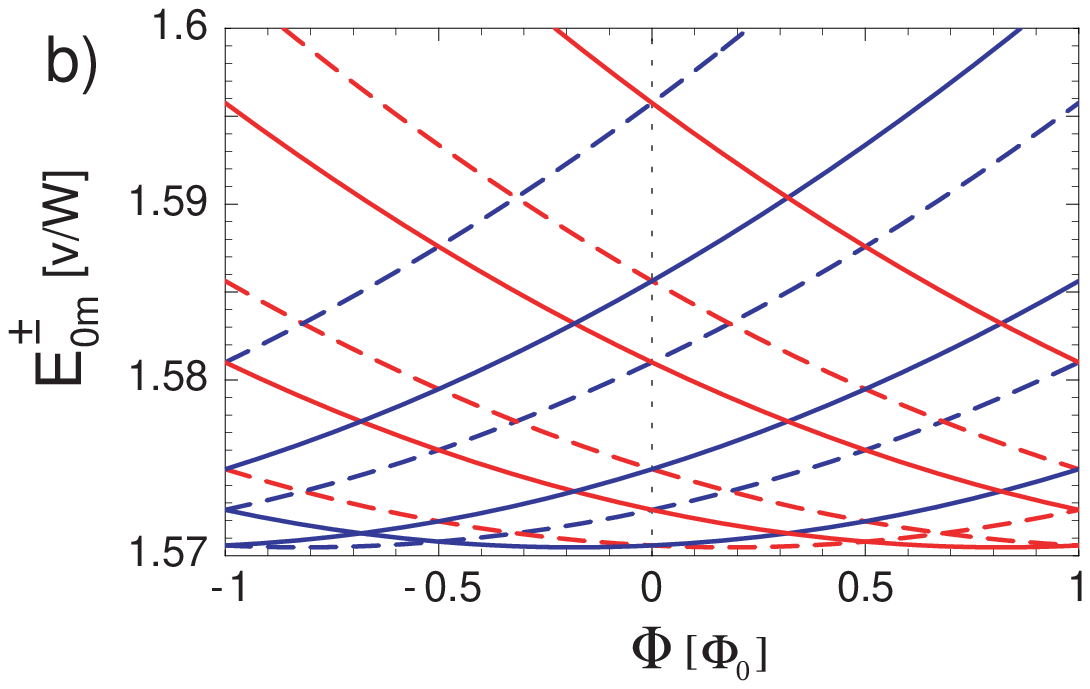}}}
\end{center}
\vspace{-0.8cm}
 \caption{Energy spectrum $E_{0m}^{\tau}$ $(E>0)$ as a function of magnetic flux $\Phi$ for various total angular-momentum values $m$ (blue $m>0$, red $m<0$)
using Eq.~(\ref{E9}) with $a/W$=10:
(a) shows a single valley only $\tau= +1$. The dotted lines are the exact numerical evaluation of
$E$ using Eq.~(\ref{5f}). In (b) we show the full spectrum
including the other valley $\tau=-1$ (dashed lines). The restoration of $\pm\Phi$ symmetry in the combined spectrum of both valleys and the lifting of
the valley degeneracy at finite $\Phi$ are clearly visible.}
\end{figure}

For $V(r)=0$,  $\chi_{1}^{\tau}(r)$ and $\chi_{2}^{\tau}(r)$ are solutions to Bessel's differential equation of order ${\overline m}-\frac{1}{2}$ and ${\overline m}+\frac{1}{2}$, respectively. Therefore, the eigenspinor for ${\widetilde
H}_{\tau}(r)$ with energy $E$ and $V(r)=0$ can be written as
\begin{equation}
\label{5e} \chi_{\tau}=
a_{\tau}\left(
\begin{array}{l}
H_{{\overline m}-\frac{1}{2}}^{(1)}(\rho)\\ i{\rm sgn}(E) H_{{\overline m}+\frac{1}{2}}^{(1)}(\rho)
\end{array}\right)
+b_{\tau}\left(
\begin{array}{l}
H_{{\overline m}-\frac{1}{2}}^{(2)}(\rho)\\ i{\rm sgn}(E)H_{{\overline m}+\frac{1}{2}}^{(2)}(\rho)
\end{array}\right),
\end{equation}
where $H_{\nu}^{(1,2)}(\rho)$ are Hankel functions of the (first, second) kind and the dimensionless radial coordinate is $\rho=|E|r/v$.
The coefficients $a_{\tau}$ and $b_{\tau}$  are determined by the boundary
condition of the ring induced by $V(r)$ (with
$V(r)\rightarrow +\infty$ outside the graphene ring). We use the infinite mass boundary condition $\psi_{\tau}=\tau({\bm n}_{\perp}\cdot {\bm \sigma})\psi_{\tau}$ where ${\bm n}_{\perp}=(-\sin\varphi,\cos\varphi)$ at $r=a+\frac{W}{2}$ and with opposite sign at
$r=a-\frac{W}{2}$. \cite{Berry,AKBE} Here $a$ is the ring radius and $W$ its width (see Fig.~1). Eliminating the coefficients $a_{\tau}$ and $b_{\tau}$ gives the
energy eigenvalue equation $z=z^{*}$ with
\begin{equation}
\label{5f} z=\frac{H_{{\overline m}-\frac{1}{2}}^{(1)}(\rho_{2})-\tau\,{\rm sgn}(E)H_{{\overline
m}+\frac{1}{2}}^{(1)}(\rho_{2})}{H_{{\overline m}-\frac{1}{2}}^{(1)}(\rho_{1})+\tau\,{\rm
sgn}(E)H_{{\overline m}+\frac{1}{2}}^{(1)}(\rho_{1})},
\end{equation}
which is equivalent to $\phi=\pi n$ with $\phi$ the phase of
$z$ and $n$ an integer. In Eq.~(\ref{5f}), we have abbreviated $\rho_{1}\equiv |E|(a-\frac{W}{2})/v$ and $\rho_{2}\equiv |E|(a+\frac{W}{2})/v$. To obtain an analytical approximation of the spectrum, we
 use the asymptotic form of the Hankel functions for
large $\rho$, including corrections up to order $1/\rho^2$.\cite{Hankeldetails} This indeed is the desired limit as $\rho=|E|r/v\sim
|E|a/v\propto a/W\gg 1$ when the ring radius is much larger than its width \cite{remarkring} and leads to the following energy eigenvalue equation
\begin{multline}
\label{E8}
 |E|=\frac{v}{W}\left(n-\tau\,\frac{{\rm
sgn}(E)}{2}\right)\pi+\frac{1}{2}\left(\frac{v}{a}\right)^2\frac{{\overline
m}^2}{|E|}\\-\frac{1}{2}\tau\,{\rm
sgn}(E)\left(\frac{v}{a}\right)^2\frac{v}{W}\frac{{\overline
m}}{|E|^2}.
\end{multline}
An iteration of Eq.~(\ref{E8}) by replacing $|E|$ on the right-hand-side
of the equation by the first (leading) term of $|E|$ gives the
energy eigenvalues (neglecting terms of ${\cal O}[(W/a)^2]$)
\begin{equation}
\label{E9}
E_{nm}^{\tau}=\pm\varepsilon_{n}\pm\lambda_{n}{\overline m}\left({\overline
m}\mp\frac{\tau}{\left(n+\frac{1}{2}\right)\pi}\right).
\end{equation}
In Eq.~(\ref{E9}), $\varepsilon_{n}=v(n+\frac{1}{2})\pi/W$, $n=0,1,2,...$, and $\lambda_{n}=\left(v/a\right)^2/2 \varepsilon_{n}$. These
energy eigenvalues are plotted as a function of flux for $n=0$ and different
values of $m$ (its half-odd integer values reflect the $\pi$-Berry phase of closed loops in graphene) in Fig.~2.
Fig.~2(a) shows the energy levels for one valley, $\tau=+1$. It is clearly visible that
$E^{\tau}(\overline{m})\neq E^{\tau}(-{\overline m})$, since effective TRS is
broken by the confinement. In Fig.~2(b), the spectrum of both valleys is shown with
full lines for $\tau=+1$, and dashed lines for $\tau=-1$. At $\Phi=0$,
$E^{\tau}(m)=E^{-\tau}(-m)$ as it should be, since real TRS is present at zero magnetic field. Crucially, however, at finite
$\Phi$, $E^{+} \neq E^{-}$ in general, showing that {\it the valley degeneracy
  is indeed lifted since effective and real TRS are broken}.
If $n\gg 1$, the term $\propto \tau \,{\overline m}$ in Eq.~(\ref{E9}) becomes
suppressed and the valley degeneracy is restored, correctly predicting that the
spectrum is insensitive to the boundary condition if $2\pi/q_{n}\ll W$, where
$q_{n}=\pi(n+\frac{1}{2})/W$ is the transverse wave number.
\begin{figure}
\begin{center}
\hbox{\resizebox{7.4cm}{!}{\includegraphics{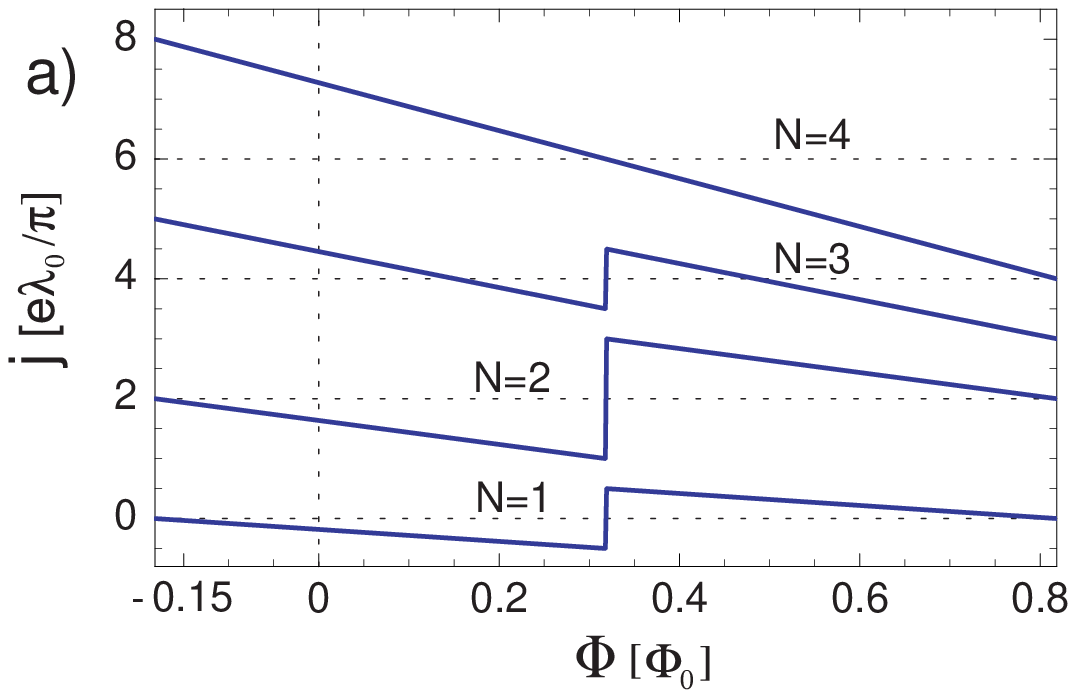}}}
\vspace{0.2cm}
\hbox{\,\,\,\,\,\,\qquad\resizebox{6.5cm}{!}{\includegraphics{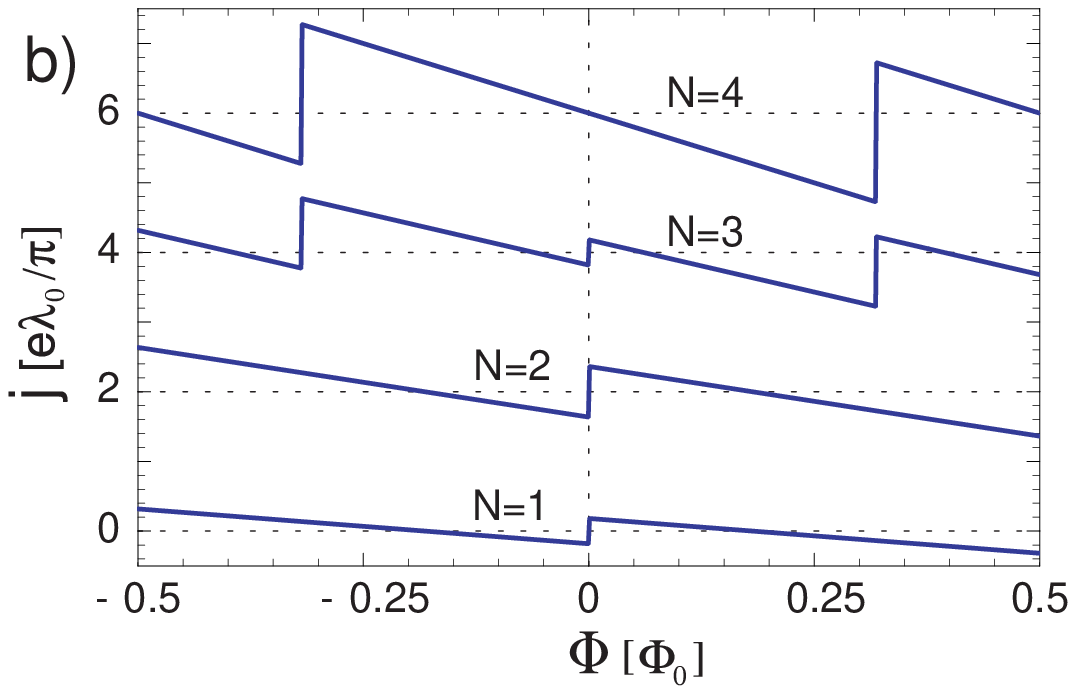}}}\end{center}
\vspace{-0.8cm}
\label{pers_fig}
\caption{ Persistent current as a function particle number $N$ (including
  spin) and flux $\Phi$ for $n=0$ ($E>0$): (a) includes only
  valley $\tau=+1$ whereas (b) includes both valleys. Curves for different $N$
  are displaced with dashed horizontal lines defining $j=0$ for each
  curve. The broken valley degeneracy is clearly visible in (b) via two
  substructures of length $\Delta\Phi=2\Phi_{0}/\pi$ and
  $\Delta\Phi=(1-(2/\pi))\Phi_{0}$ whereas  (a) predicts a non-zero persistent
  current at $\Phi=0$ due to effective TRS breaking in a single valley.}
\end{figure}
We show next that a
broken valley degeneracy results in observable features in the persistent
current and the conductance through the ring.

\subsection{Persistent current and conductance}
The persistent
current in the closed ring is given at zero temperature by
$j=-\sum_{\tau}\sum_{nm}\partial E_{nm}^{\tau}/\partial \Phi$ where the sum
runs over all occupied states. In Fig.~3,  we show the
persistent current as a function of number of electrons on the ring $N$
(including spin) and magnetic flux relative to the half-filled band. (We
subtract the contribution to the persistent current that arises from all
states with $E<0$.) The persistent current is periodic in $\Phi$
with period $\Phi_{0}$. In Fig.~3(a), only one valley, $\tau=+1$ is considered and a {\it finite} persistent current at $\Phi=0$ is predicted. Therefore, a non-zero persistent
current at zero flux detects valley polarization. In Fig.~3(b), we show the case of equal
population of both valleys. Then, the persistent current as a function of flux is zero at $\Phi=0$, but
shows a substructure (kinks at $\Phi\neq 0$) within one period directly
related to the broken valley degeneracy at finite flux. 
%%%%%%%%%%%%%%%%%%%%%%%%%%%%%%%%%%%%%%%%%%%%%%%%%%%%%%%%%%%%%%%%%%%
We note that this substructure is due to the linear term in $\overline{m}$ of the spectrum Eq.~(\ref{E9}) which is prominent within the first few transverse modes $n$ which can host many electrons $N$. 
%%%%%%%%%%%%%%%%%%%%%%%%%%%%%%%%%%%%%%%%%%%%%%%%%%%%%%%%%%%%%%%%%%%%%

In Fig.~4, we plot the conductance through the ring weakly coupled to leads as a function of Fermi energy $E_{F}$ (or gate
voltage) assuming a constant interaction model \cite{Leobook} with charging
energy $U$.\cite{charging}
\begin{figure}[t]
\begin{center}
\hbox{\qquad\resizebox{6.8cm}{!}{\includegraphics{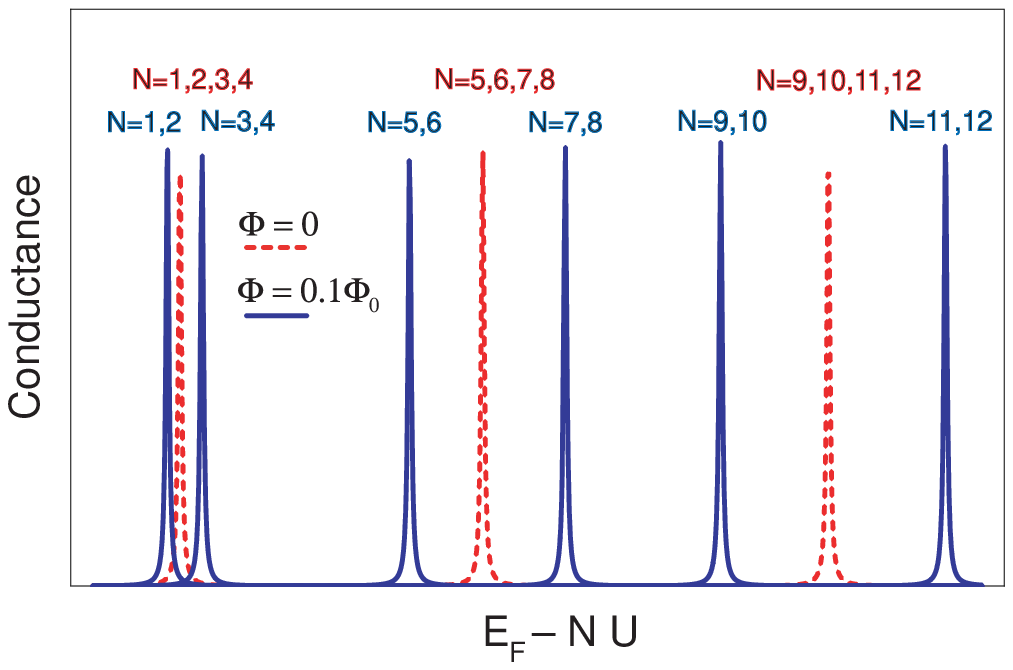}}}
\end{center}
\vspace{-0.8cm}
\caption{Ring conductance assuming a constant interaction model with
charging energy $U$ for the first 12 electrons in the conduction band $(E>0)$: At $\Phi=0$ (dashed), the
conductance shows a four-fold symmetry as a function of Fermi energy $E_{F}$ in the leads
due to  spin- and valley-degeneracy. At finite magnetic flux
($\Phi/\Phi_{0}=0.1$, full line), the conductance peaks shift due
to breaking of the valley degeneracy. Each peak is labelled with the filling factor $N$ at this specific resonance.}
\end{figure}
At $\Phi=0$, the conductance exhibits a four-fold
symmetry due to spin and valley
degeneracy. A finite flux breaks the valley degeneracy
which is observable via a splitting
of the conductance peaks moving with magnetic flux, see Fig.~4.

\subsection{Valley qubit}
We now turn to the question of how to make use of the broken valley
degeneracy in order to directly address the valley degree of freedom in
graphene experimentally (valleytronics \cite{valley-filter}).
The valley
degree of freedom forms (in principle)
a two-level system that can be represented
by an isospin $|+\rangle$ for valley $\tau=+1$ and
$|-\rangle$ for valley $\tau=-1$.
We point out that the graphene ring
 weakly coupled to current leads could be used
 to investigate the {\it relaxation} and {\it coherence} of such valley isospins.
 \begin{figure}[h]
\begin{center}
\hbox{\resizebox{7.4cm}{!}{\includegraphics{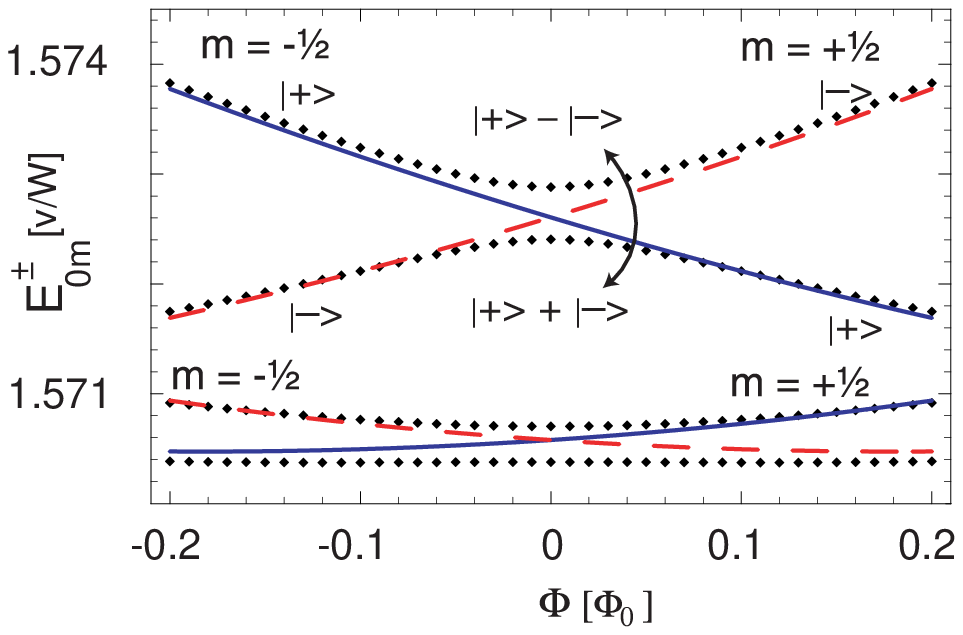}}}
%\quad\resizebox{7.0cm}{!}{\includegraphics{}}}
\end{center}
\vspace{-1cm}
\caption{Valley qubit. The lowest four levels (for $n=0$ and $a/W$=10) are
  shown as a function of flux. Blue and red (dashed) lines correspond to valleys
  $\tau=+1$ and $\tau=-1$, respectively. The dotted lines take into account
  small level mixing leading to anticrossings. The flux $\Phi$ is used to
  switch from $|+\rangle$ to the crossing point with the new eigenstates being
  superpositions of $|+\rangle$ and $|-\rangle$ as indicated in the figure. }
\end{figure}
Close to a degeneracy point of two levels belonging to different valleys
(e.g. at $\Phi=0$), Eq.~(\ref{E9}) predicts a {\it valley splitting} of states
with fixed $m$-values controllable by flux, similar to the Zeeman-splitting
for electron spins in a magnetic field. In semiconductor quantum dots,
such pairs of
spin-split states can be addressed via electron tunneling from/to leads weakly
coupled to the quantum dot and can be used for read-out of single spins
\cite{spin-read-out} or measuring their relaxation ($T_{1}$) time.\cite{T1} The graphene ring could be used in very much the same way to
measure the intrinsic valley isospin relaxation time $T_{1}$ in graphene as well as the valley isospin polarization.

In Fig.~5, we show the situation when some small level-mixing leads to avoided
crossings of valley-split states near the degeneracy point $\Phi=0$. Such
valley mixing naturally appears through boundary roughness of the ring or
atomic defects in the bulk. 
Using the magnetic flux as a knob, we can sweep the
system from a $|+\rangle$ groundstate level, filled with one electron, to a superposition
$(|+\rangle+|-\rangle)/\sqrt{2}$ and further to a $|-\rangle$ groundstate, see
Fig.~5. Such a situation can be used to produce Rabi oscillations of the
valley isospin states by
tuning the system fast (non-adiabatically) from $|+\rangle$ to the degeneracy
point where the spin will oscillate between $|+\rangle$ and $|-\rangle$ in
time: $\cos(\Delta t)|+\rangle-i\sin(\Delta t)|-\rangle$, where $2\Delta$ is
the energy splitting at the degeneracy point.\cite{qubitH}

We expect that this qubit is rather robust if intervalley scattering is weak, since time-reversal symmetry assures the (approximate) degeneracy of states from different valleys at zero flux $\Phi$. Indeed, the Hamiltonian Eq.~(1) has the same spectrum in both valleys at zero flux, independent on the shape of the mass potential $V(x,y)$. \cite{spectrumproof} This means that we do not rely on a special symmetry of the confining potential (like the circle discussed here). In addition, long-range disorder will also not lift the valley degeneracy.

%%%%%%%%%%%%%%%%%%%%%%%%%%%%%%%%%%%%%%%%%%%%%%%%%%%%%%%%%%%%%%%%%%%%%%
\subsection{Valley isospin-orbit coupling}
%%%%%%%%%%%%%%%%%%%%%%%%%%%%%%%%%%%%%%%%%%%%%%%%%%%%%%%%%%%%%%%%%%%%%%
In an open ring geometry with adiabatic contacts to leads, new interesting
coherent rotations of the valley isospin
occur while propagating along the ring. The linear term in ${\overline m}$
in Eq.~(\ref{E9}) can be thought of as a {\it valley isospin-orbit
  coupling term}, since the valley isospin $\tau$ couples to the orbital motion
${\overline m}$. A general incoming spinor is a superposition of spinors
belonging to different valleys. Due to the valley isospin-orbit coupling, the
angular momentum $m$ [determined by the incoming (continuous) energy $E$ and
the applied magnetic flux $\Phi$ via Eq.~(\ref{E9})] will be different for the
two valleys. Consequently,  the spinor in Eq.~(\ref{E3}) will pick up
different phases $\exp(im\varphi)$ for the two valleys while propagating along
the ring thereby rotating the
valley isospin in a transport experiment.

%%%%%%%%%%%%%%%%%%%%%%%%%%%%%%%%%%%%%%%%%%%%%%%%%%%%%%%%%%%%%%%%%%%%%%
%%%%%%%%%%%%%%%%%%%%%%%%%%  NEW SECTION %%%%%%%%%%%%%%%%%%%%%%%%%%%%%
\section{Spectrum for a hexagonal ring with zigzag edges}
%%%%%%%%%%%%%%%%%%%%%%%%%%%%%%%%%%%%%%%%%%%%%%%%%%%%%%%%%%%%%%%%%%%%%%
Here, we compare our analytical model with the infinite mass boundary described in Section II, to a ring with strong intervalley scattering. We numerically investigate the spectrum of a ring of hexagonal form with zigzag 
edges as shown in Fig.~6. (Electrical conduction through this geometry was studied in Ref.\
\onlinecite{Adamringpaper}.)
%%%%%%%%%%%%%%%%%%%%%%%%%%%%%%%%%%%%%%%%%%%%%%%%%%%%%%%%%%%%%%%%%%%%%
\begin{figure}[h]
\begin{center}
\hbox{\,\,\qquad\qquad\resizebox{4.8cm}{!}{\includegraphics{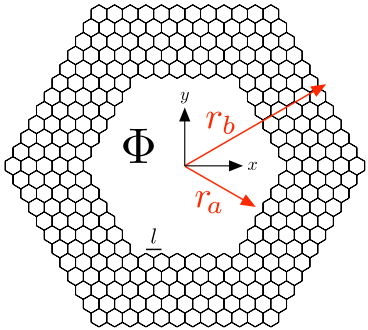}}}
\end{center}
\vspace{-0.8cm}
\caption{A hexagonal ring with zigzag edges of inner radius $r_a$ and outer radius $r_b$ and flux through the hole $\Phi$. Here, $r_a=3\sqrt{3}l$ and $r_b=6\sqrt{3}l$ with $l$ the lattice spacing}
\end{figure}
%\vspace{0.8cm}
%\hbox{\,\,\,\,\qquad\resizebox{5.2cm}{!}{\includegraphics{valleyarrows3}}}
\begin{figure}
\begin{center}
\hbox{\resizebox{8cm}{!}{\includegraphics{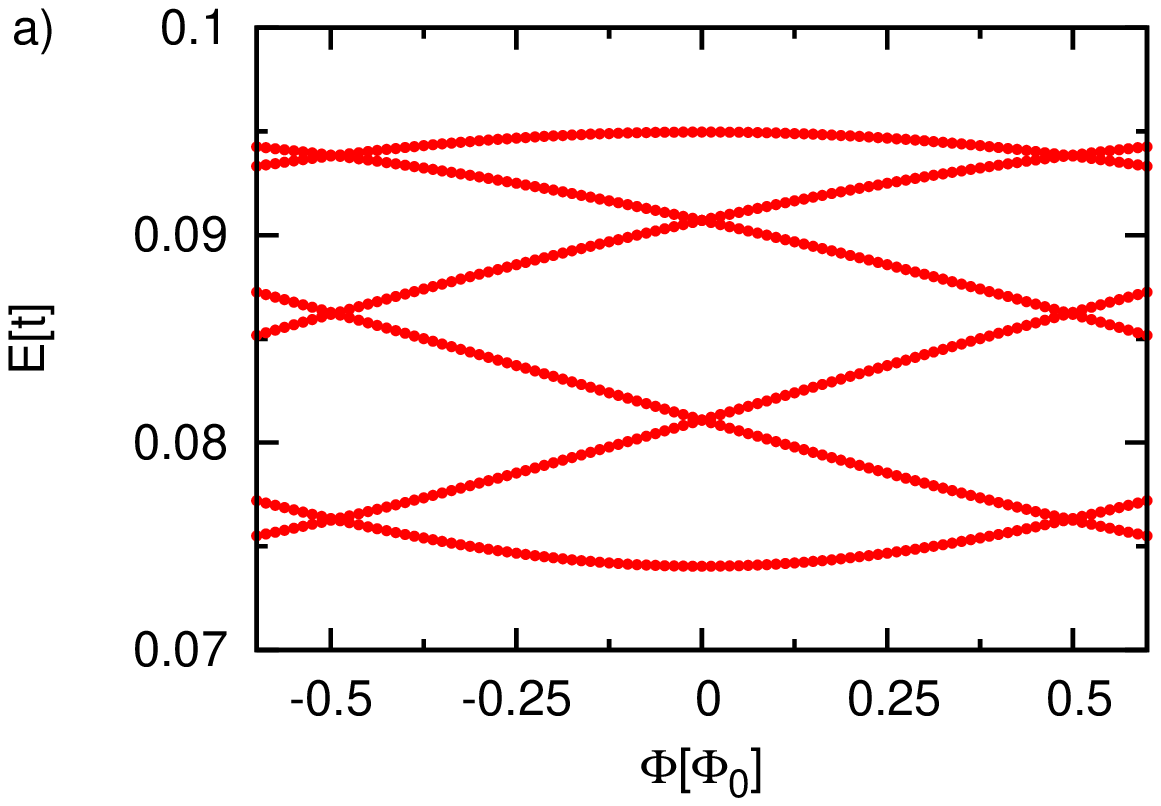}}}

\hbox{\,\,\resizebox{7.2cm}{!}{\includegraphics{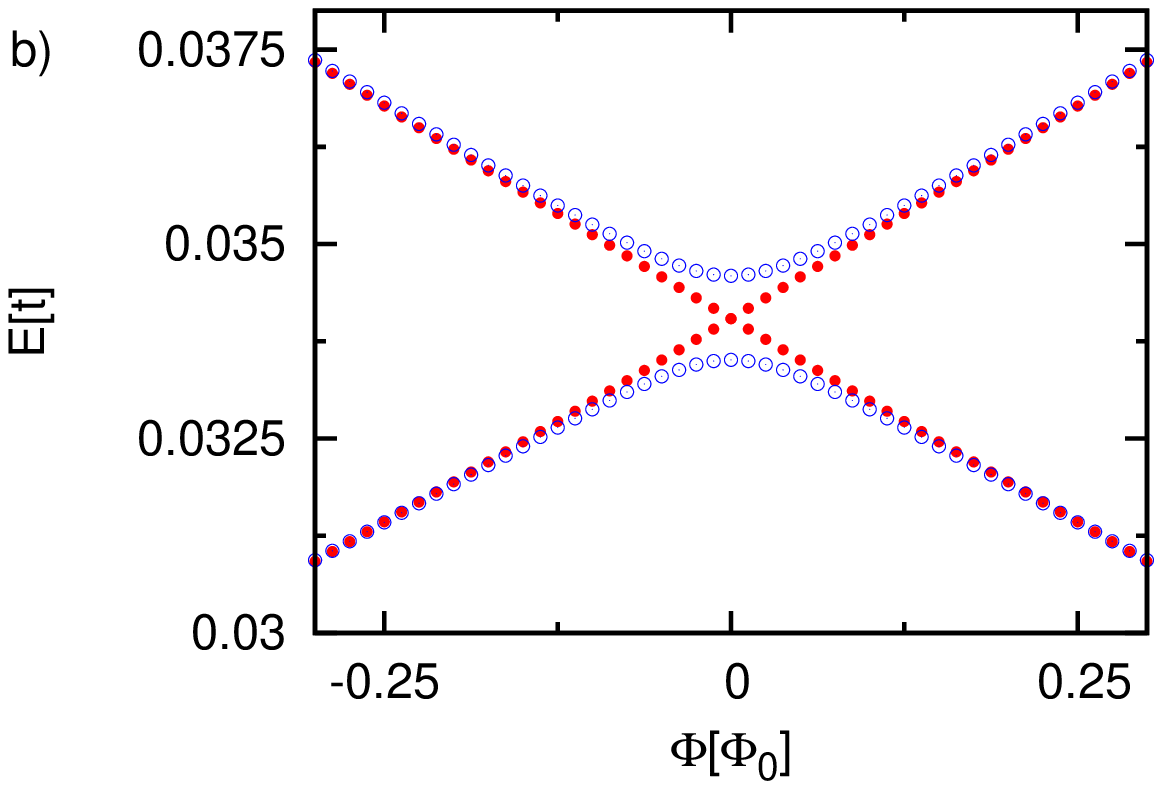}}}
\end{center}
\vspace{-0.8cm}
\caption{Plot (a) shows an energy band of the hexagonal ring (see Fig.~6) with ring dimensions $r_{a}=7\sqrt{3}l$ and $r_{b}=14\sqrt{3}l$ in the lowest mode $0<E\lesssim 0.34 t$. \cite{modewidth} The levels are grouped into bands containing 6 levels. The top most and the lowest level in each band is non-degenerate whereas the middle four levels are two-fold degenerate at zero flux. This degeneracy is lifted by the flux through the ring. In (b) we contrast the perfect crossing of two levels at zero flux (red dots) with anticrossed levels (blue circles) induced by the addition of one unit cell to each of two parallel arms of the ring  [we have shifted the energy axis for the asymmetric case (blue circles) by $ +3\cdot 10^{-3} t$ for better comparison].}
\end{figure}
%%%%%%%%%%%%%%%%%%%%%%%%%%%%%%%%%%%%%%%%%%%%%%%%%%%%%%%%%%%%%%%%%%%%
In a zigzag nanoribbon, the valley isospin is a good quantum number, i.e. the zigzag boundary does not mix valleys.\cite{BF} Since two neighboring arms of the ring are rotated by $60^{\circ}$ with respect to each other, the roles of the A and B sublattices are interchanged in subsequent segments. Explicitly, this means, that if a zigzag edge is terminated on a A side, it will be terminated on a B side at a neighboring arm of the ring. Equivalently, in the reciprocal ($\mathbf{k}$-) space, this implies that equivalent states of subsequent zigzag nanoribbon segments are lying in opposite valleys.
This necessarily induces intervalley mixing at the corners between two subsequent zigzag nanoribbon segments. This mixing is very strong in the lowest mode of the ring, where the direction of motion and the valley is tightly coupled in each arm of the hexagonal ring \cite{BF,valley-filter} (the zigzag edge is therefore another example where effective TRS in a single valley is broken). An electron wave, approaching a corner of the hexagon in one valley, is either transmitted into the next arm, or reflected back into the same arm. In both cases, the valley index has to flip. 

We investigate the spectrum of such a ring numerically in a tight-binding approach with Hamiltonian
\begin{equation}
H=\sum\limits_{i,j}t_{ij}|i\rangle\langle j|+\sum\limits_{i}\epsilon_{i}|i\rangle\langle i|.
\end{equation}
The hopping element in the presence of a magnetic flux is $t_{ij}=-t\exp[-i(2\pi/\Phi_{0})\int_{{\bf r}_{j}}^{{\bf r}_i}d{\bf r}\cdot{\bf A}]$ where $\bf {A}$ is the vector potential and $\epsilon_{i}=0$ are the on-site energies. 
The vector potential is chosen as ${\bf A}=(A_x,0,0)$ with
\begin{equation}
 A_x=B\left[y_{a}\Theta(y)-y_{a}\Theta(-y)\right]\times
\Theta(L_C/2-|x|), 
\end{equation}
where $y_{a}(x)=\min[r_{a},\sqrt{3}(L_{C}/2-|x|)]$, with
$L_C=4r_a/\sqrt{3}$ for the perfect hexagon (shown in Fig.~6) and $L_C=4r_a/\sqrt{3}+l$ for a hexagon ring with one unit cell added to the top and bottom arm. This represents a uniform magnetic field $B$ inside the ring hole and zero outside.
The spectrum as a function of magnetic flux $\Phi=2\sqrt{3}r_{a}^2 B$  is shown in Fig.~7 in an energy window which lies well within the lowest mode of a zigzag nanoribbon of width $W=r_{b}-r_{a}$. \cite{modewidth} 
A band of levels in the lowest mode is shown in Fig.~7(a). Within that mode,
the spectrum follows a clear pattern which is observed for generic values of
$r_a$ and $r_b$. It consists of bands separated by energy gaps. Each band
hosts six levels. The top and bottom level is non-degenerate with $dE/d\Phi=0$
at zero flux (corresponding to standing waves). The other four levels are
two-fold degenerate at $\Phi=0$ with a broken degeneracy at finite flux. These
levels correspond to right and left-going states in the ring. 

We remark that this level pattern reflects the scattering off a periodic array
of six scatterers subjected to periodic boundary
conditions. \cite{periodic_array} It is to be noted that the orbital
degeneracy of the hexagonal ring can be tuned by the flux, 
 similar to the ring with the smooth confinement discussed in Section II.  
If the 6-fold rotational symmetry of the ring is broken, the crossings at zero
flux become slightly avoided as is shown in Fig.~7(b) (blue circles) where we
have added one unit cell to two of the parallel arms of the ring (this corresponds to a length change of the arms by about 5$\%$). This shows
that our results are also relevant for rings with a sligthly reduced
symmetry. Note that the sensitivity of the level crossing at zero flux to the ring geometry is consistent with strong intervalley scattering where time-reversal symmetry does not protect the degeneracy at zero flux.

\section{Conclusion}
We have analyzed the Aharonov-Bohm effect in graphene rings. We have
investigated two different ring systems -- a ring with a smooth boundary
(with zero or weak intervalley scattering) and a hexagonal ring with
zigzag edges. For the ring with a smooth boundary, the combined effect
of the effective time reversal symmetry (TRS) breaking within a single
valley induced by a smooth boundary and the applied magnetic flux -- breaking
the real TRS -- gives us at hand a controllable tool to break the valley
degeneracy in such rings. We have shown that this effect of a broken valley degeneracy by flux is revealed in the persistent current and in the ring conductance. This tool could be useful 
for spin-based or valley-based quantum computing. The presence of a degenerate pair of levels from different valleys at zero flux is assured by time-reversal symmetry in the absence of intervalley scattering. Therefore, the proposed effect is not sensitive to the actual geometry of the ring.

We have also considered the opposite case of strong intervalley scattering by investigating numerically a hexagonal ring with lattice-terminated zigzag edges. Here, strong intervalley scattering is induced by the corners of the ring at low energies. We found that
the orbital degeneracy of graphene can still be tuned by the flux similar to
the ring with a smooth boundary. This effect, however, relies on a certain degree of symmetry of the ring as we show by sligthly distorting the ring. 
We therefore conclude that the orbital degeneracy in graphene rings  can be controlled with an Aharonov-Bohm flux in rings with zero or weak intervalley scattering and in systems with strong intervalley scattering if the ring possesses an (approximate)
geometric symmetry.

We acknowledge helpful discussions with A.R. Akhmerov and J.H. Bardarson. This
work was financially supported by the Dutch Science Foundation NWO/FOM, the
Swiss NSF, and the NCCR Nanoscience. A. Rycerz acknowledges support by the Polish Ministry of Science (Grant No.
1-P03B-001-29) and by the Polish Science Foundation (FNP).

%%%%%%%%%%%%%%%%%%%%%%%%%%%%%%%%%%%%%%%%%%%%%%%%%%%%%%%%%%%%%%%%%%%%%


\begin{references}
%%%%%%%%%%%%%%%%%%%%%%%%%%%%%%%%%%%%%%%%%%%%%%%%%%%%%%%%%%%%%%%%%%%%%

\bibitem{Geimreview}
For a recent review on the topic, see
A.K. Geim and K.S. Novoselov, Nature Materials {\bf 6}, 183 (2007).

\bibitem{AMFG}
S.V. Iordanskii and A.E. Koshelev, JETP Letters {\bf 41}, 574 (1985); S.V. Morozov {\it et al.}, Phys. Rev. Lett. {\bf 97}, 016801 (2006);
A.F. Morpurgo and F. Guinea, Phys. Rev. Lett. {\bf 97}, 166804 (2006).

\bibitem{WAL}
H. Suzuura and T. Ando, Phys. Rev. Lett. {\bf 89}, 266603 (2002); D.V. Khveshchenko, Phys. Rev. Lett. {\bf 97}, 036802 (2006); E. McCann, K. Kechedzhi, V.I. Fal'ko, H. Suzuura, T. Ando, and B.L. Altshuler, Phys. Rev. Lett. {\bf 97}, 146805 (2006).

\bibitem{graphenespinqubits}
B. Trauzettel, D.V. Bulaev, D. Loss, and G. Burkard, Nature Physics {\bf 3}, 192 (2007).

%\bibitem{AB}
%Y. Aharonov and D. Bohm, Phys. Rev. {\bf 115}, 485 (1959).

\bibitem{Berry}
M.V. Berry and R.J. Mondragon, Proc. R. Soc. Lond. A{\bf 412},
53 (1987).

\bibitem{Antonremark}
Any boundary condition that does not mix the valleys can be written as
$\psi={\cal M} \psi$, with ${\cal M}=\bm{n}_{\perp}\cdot\bm{\sigma}$ determined by a unit
vector $\bm{n}_{\perp}$ in the plane tangent to the boundary.\cite{AKBE} It holds that $[{\cal
  M},{\cal T}]\neq 0$ where ${\cal T}=i\sigma_{y}{\cal C}$ (${\cal C}$ denotes complex conjugation) is
the time reversal operation in a single valley.

\bibitem{AKBE} A.R. Akhmerov and C.W.J. Beenakker, Phys. Rev. Lett. {\bf 98}, 157003 (2007).

\bibitem{Giova07}
 G. Giovannetti, P.A. Khomyakov, G. Brocks, P.J. Kelly, and J. van den Brink,
Phys. Rev. B {\bf 76}, 073103 (2007).

\bibitem{Zhou07}
S.Y. Zhou, G.-H. Gweon, A.V. Fedorov, P.N. First, W.A. de Heer, D.-H. Lee,
F. Guinea, A.H. Castro Neto, and A. Lanzara, arXiv:0709.1706 (2007). 

\bibitem{Meijer}
F.E. Meijer, A.F. Morpurgo, and T.M. Klapwijk, Phys. Rev. B  {\bf
66}, 033107 (2002).

\bibitem{Frustaglia}
D. Frustaglia and K. Richter, Phys. Rev. B {\bf 69}, 235310 (2004).

\bibitem{Hankeldetails}
We use
$H_{\nu}^{(1)}(\rho)=\sqrt{2/(\pi\rho)}\exp[i(\rho-\nu\pi/2-\pi/4)]\{1+\delta_{\nu}(\rho)\}$
with
$\delta_{\nu}(\rho)=-(4\nu^2-1)(4\nu^2-9)/128\rho^2+i(\nu^2-1/4)/2\rho+{\cal
  O}(\rho^{-3})$ and $H_{\nu}^{(2)}(\rho)=[H_{\nu}^{(1)}(\rho)]^{*}$.

\bibitem{remarkring}
In the regime $a\sim W$, the breaking of the valley degeneracy with flux is also observed as we find numerically using Eq.~(\ref{5f}).

\bibitem{Leobook}
L.P. Kouwenhoven, C.M. Marcus, P.L. McEuen, S. Tarucha, R.M. Westervelt, and N.S. Wingreen, in {\it Mesoscopic Electron Transport}, NATO ASI
Series E, Vol. 345 (Kluwer Academic Publishers, Dordrecht, 1997) ed. by L.L. Sohn, L.P. Kouwenhoven, and G. Sch\"on, p. 105.

\bibitem{charging}
We estimate $U\sim 4$ meV for $a=0.5$ $\mu$m and $a/W=5$ with a simple ring capacitor model with plate separation of
$285$ nm and ${\rm SiO}_{2}$ dielectric. The single-particle levelspacing for $n=0$ and small ${\overline m}$ is $\sim \lambda_{0}=(v/a)(W/a)/\pi\sim 66$ $\mu$eV and grows linearly with ${\overline m}$.

\bibitem{valley-filter}
A. Rycerz, J. Tworzyd\l o, and C.W.J. Beenakker, Nature Physics {\bf 3}, 172
(2007).

\bibitem{spin-read-out}
J.M. Elzerman, R. Hanson, L.H.W. van Beveren, B. Witkamp, L.M.K. Vandersypen, and L.P. Kouwenhoven, Nature {\bf 430}, 431 (2004).

\bibitem{T1}
R. Hanson, B. Witkamp, L.M.K. Vandersypen, L.H.W. van Beveren, J.M. Elzerman, and L.P. Kouwenhoven, Phys. Rev. Lett. {\bf 91}, 196802 (2003).

%\bibitem{functionbook}
%N.N. Lebedev in {\it Special Functions And Their Applications}.

\bibitem{qubitH}
Near $\Phi=0$, the spectrum is approximated by a ``valley qubit''-Hamiltonian
${\cal
  H}_{0m}=2\lambda_{0}(\Phi/\Phi_{0})[m-(1/\pi)]\tau_{z}-\Delta\tau_{x}+c(m)$
with $c(m)$ being independent of $\Phi$, and $\tau_{z}$ and $\tau_{x}$
act in valley space.
\bibitem{spectrumproof}
This is shown as follows: If $\psi_{\tau}$ is an eigenstate of $H_{\tau}$, then ${\cal T}\psi_{\tau}$ with ${\cal T}=i\sigma_y{\cal C}$  the time-reversal operation in a single valley, is an (orthogonal) eigenstate of $H_{\tau}$ with mass potential $-V(x,y)$ and with the same energy. \cite{Berry} But $H_{\tau}$ with mass potential $-V(x,y)$ is identical to $H_{-\tau}$ with mass potential $V(x,y)$. 
\bibitem{Adamringpaper}
A. Rycerz and C.W.J. Beenakker, arXiv:0709.3397.
\bibitem{BF}
L. Brey and H.A. Fertig, Phys. Rev. B {\bf 73}, 235411 (2006).
\bibitem{modewidth}
For $W\gg l$, the energy spacing $\delta$ between the 1st (lowest) and 2nd mode in a zigzag nanoribbon is $\delta=(3/2)\Delta$ with $\Delta=(1/2)\sqrt{3}\pi t l/W$. \cite{valley-filter} For the ring dimensions used in Fig.~7, this gives $\delta\sim 0.34 t$.
\bibitem{periodic_array}
R. Gilmore, {\it Elementary Quantum Mechanics In One Dimension} (The Johns Hopkins University Press, Baltimore, 2004), Chaps.~37 and 38.
\end{references}
\end{document}